\documentclass[english,aps]{revtex4}
\usepackage[T1]{fontenc}
\usepackage{graphicx}
\usepackage{amssymb}

\makeatletter

\usepackage{bm}

\usepackage{babel}
\makeatother

\begin{document}

\title{BCS-BEC crossover in a relativistic boson-fermion model beyond mean
field approximation}

\author{Jian Deng, Jin-cheng Wang and Qun Wang }

\affiliation{Interdisciplinary Center for Theoretical Study and Department of
Modern Physics, University of Science and Technology of China, Anhui
230026, People's Republic of China}

\begin{abstract}
We investigate the fluctuation effect of the di-fermion field in the
crossover from Bardeen-Cooper-Schrieffer (BCS) pairing to a Bose-Einstein
condensate (BEC) in a relativistic superfluid. We work within the
boson-fermion model obeying a global U(1) symmetry. To go beyond the
mean field approximation we use Cornwall-Jackiw-Tomboulis (CJT) formalism
to include higher order contributions. The quantum fluctuations of
the pairing condensate is provided by bosons in non-zero modes, whose
interaction with fermions gives the two-particle-irreducible (2PI)
effective potential. It changes the crossover property in the BEC
regime. With the fluctuations the superfluid phase transition becomes
the first order in grand canonical ensemble. We calculate the condensate,
the critical temperature $T_{c}$ and particle abundances as functions
of crossover parameter the boson mass. 
\end{abstract}
\maketitle

\section{Introduction}

Fermion pairings are mechanisms for superconductivity and superfluidity.
In the case of weak attractive interaction between fermions the pairings
are well described in Bardeen-Cooper-Schrieffer (BCS) theory \cite{bcs},
where fermion pairs are typically of a size much larger than the mean
interparticle distance. In some sense the pairing can be regarded
as taking place in momentum space. As the attractive interaction gets
strong enough fermion pairs become real bosonic bound states. Below
a critical temperature macroscopically large number of these molecular
bosons occupy the ground state and form a Bose-Einstein condensate
(BEC). The idea of covering BEC regime in an extended BCS theory was
proposed by Eagles \cite{Eagles:1969} and Leggett \cite{Leggett:1980}.
A quantitative description of the crossover at finite temperatures
from weak to strong coupling regime was given by Nozieres and Schmitt-Rink
\cite{Nozieres:1985}, who wrote down the universal pair wave function
which can be reduced to the correct ground states at BCS and BEC limit.
The BCS-BEC crossover have been extensively investigated in cold atom
systems since such a crossover can be observed in experiments by tuning
the magnetic field around Feschbach resonance to achieve situations
with different scattering length among fermionic atoms \cite{regal:2004,Bartenstein:2004,Zwierlein:2004,Kinast:2004,Bourdel:2004}. 

Recently there is a growing interest in extending the theory of the
BCS-BEC crossover to relativistic systems. The crossover from pion
condensation to Cooper pairing of quarks and antiquarks at large isospin
densities provides such an example \cite{Son:2000xc,Son:2000by,He:2005nk,Zablocki:2008sj}.
Another example is color superconductivity where quarks form Cooper
pairs on the Fermi surface due to an attractive interaction mediated
by gluon exchange \cite{Collins:1974ky,Barrois:1977xd,Bailin:1983bm,Alford:1997zt,Rapp:1997zu,Alford:1998mk,Son:1998uk,Pisarski:1999tv,Hong:1999fh}
(for reviews, see, for instance, \cite{Rajagopal:2000wf,Alford:2001dt,Schafer:2003vz,Casalbuoni:2003wh,Rischke:2003mt,Buballa:2003qv,Shovkovy:2004me,Alford:2007xm}).
Because of asymptotic freedom, color superconductivity at very large
densities can be studied in a weak-coupling approach \cite{Son:1998uk,Pisarski:1999tv,Brown:1999aq,Hong:1999fh,Wang:2001aq,Schmitt:2002sc,Reuter:2004kk,Wang:2004pa}
using perturbative QCD (see, for example, \cite{Braaten:1989mz,Blaizot:2003tw,Kraemmer:2003gd}).
For moderate densities more phenomenological models such as the Nambu-Jona-Lasinio
(NJL) model \cite{Buballa:2003qv,Hatsuda:1994pi} or self-consistent
Dyson-Schwinger equations \cite{Nickel:2006vf,Marhauser:2006hy}
are needed. All these approaches usually assume BCS-like picture.
When the coupling strength turns even stronger, the BEC-like diquark
pairing may set in. A lot of authors used the NJL model to study the
BCS-BEC crossover in relativistic superconductors or superfluids \cite{Nishida:2005ds,Abuki:2006dv,Sun:2007fc,He:2007yj,Kitazawa:2007im,Kitazawa:2007zs,Brauner:2008td}.
In our previous work we set up a theory for the crossover in relativistic
superfluids which includes both bosonic and fermionic degrees of freedom
\cite{Deng:2006ed}. It is inspired by the boson-fermion model of
superconductivity \cite{Ranninger:1985,Friedberg:1989gj,Friedberg:1990eg}.
We only addressed the mean-field approximation of the model. The crossover
is realized by tuning the difference between the boson mass and boson
chemical potential. 

In this paper we extend our previous work \cite{Deng:2006ed} by
systematic incorporating the fluctuation effect in the crossover from
propagating modes of the di-fermion field within Cornwall-Jackiw-Tomboulis
formalism (CJT) \cite{Cornwall:1974vz}. Our starting point is the
boson-fermion model with a global U(1) symmetry. Similar efforts have
been made in the NJL model \cite{He:2007yj,Brauner:2008td}. The
CJT formalism has been employed to analyze fluctuation effects in
color superconductors \cite{Giannakis:2004xt,Noronha:2006cz}. The
inclusion of fluctuation contributions leads to the change of the
second order phase transition of normal to superfluid to a first order
one. The paper is organized as follows. In Sec. II we re-write the
boson-fermion model in terms of the Higgs and Nambu-Goldstone fields.
In Sec. III the effective potential is evaluated up to two-loops including
the mean field as well as the fluctuation contributions. The Dyson-Schwinger
(DS) equations are derived from the effective potential in Sec. IV.
The gap and charge density equations are obtained in Sec. V by taking
derivatives of the effective potential with respect to the diquark
condensate and the chemical potential respectively. The numerical
results are presented and analyzed in Sec. VI. Finally we discuss
about our results and draw some conclusions in the last section. 

Our convention for the metric tensor is $g^{\mu\nu}=\textrm{diag}(1,-1,-1,-1)$. Our units are $\hbar=c=k_{B}=1$. Four-vectors are denoted by capital letters, $K\equiv K^{\mu}=(k_{0},\mathbf{k})$ with $k=|\mathbf{k}|$. Fermionic Matsubara frequencies are $\omega_n=ik_0=(2n+1)\pi T$, while bosonic ones are $\omega_n=ik_0=2n\pi T$ with the temperature $T$ and $n$ an integer. Our convention for the Feynman rules in finite temperature and density field theory are chosen as compared to those at zero temperature and density, where the propagator and the vertex for a quantum field $\phi (X)$ ($X$ is a real time-space coordinate) in the theory are like, e. g. $iG(X_1,X_2)=\left<T\phi  (X_1)\phi  (X_2)\right>$ and $i\Gamma = i\delta ^n/[\delta \phi (X_1) \delta \phi (X_2) \cdots \delta \phi (X_n)]$. At finite temperature and density, we replace $iG\rightarrow -G$ and $i\Gamma \rightarrow \Gamma$. The energy unit is chosen to be the fermion momentum cutoff $\Lambda$ appearing in fermion momentum integrals.

\section{Boson-fermion model}

In order to describe the fluctuation effects from the propagating
modes of di-fermion bosons, we start from the following Lagrangian
which respects the global U(1) symmetry corresponding to total particle
number conservation, \begin{eqnarray}
\mathcal{L}(\Phi,\Psi) & = & -\frac{1}{2}\overline{\Psi}S_{0}^{-1}\Psi+\left|(\partial_{\nu}-i\mu_{b}\delta_{\nu0})\Phi\right|^{2}-m_{b}^{2}|\Phi|^{2}\nonumber \\
 &  & +\frac{1}{2}\left(\Phi^{\dagger}\overline{\Psi}\widehat{\Gamma}\Psi-\Phi\overline{\Psi}\widehat{\Gamma}^{\dagger}\Psi\right).\label{eq:lag01}\end{eqnarray}
In the first term or the fermion sector, $\Psi=\left(\begin{array}{c}
\psi\\
\psi_{C}\end{array}\right)$ and $\overline{\Psi}=(\overline{\psi},\overline{\psi}_{C})$ are
spinors for fermions in the Nambu-Gorkov (NG) basis, where the charge
conjugate spinor is defined by $\psi_{C}=C\overline{\psi}^{T}$ and
$\overline{\psi}_{C}=\psi^{T}C$ with $C=i\gamma^{2}\gamma^{0}$.
In momentum space the inverse fermion propagator $S_{0}^{-1}$ is
given by \begin{equation}
S_{0}^{-1}(P)=-\left(\begin{array}{cc}
\gamma_{\mu}P^{\mu}+\mu\gamma^{0}-m & 0\\
0 & \gamma_{\mu}P^{\mu}-\mu\gamma^{0}-m\end{array}\right),\label{eq:fermion-prop01}\end{equation}
where $\mu$ and $m$ are the chemical potential and the mass of fermions
respectively. In the second and third terms of the Lagrangian (\ref{eq:lag01})
the boson chemical potential and the mass are denoted by $\mu_{b}=2\mu$
and $m_{b}$. In the boson-fermion interaction part, the last term
of the Lagrangian (\ref{eq:lag01}), we have defined the interaction
vertices \begin{equation}
\widehat{\Gamma}=2gi\gamma_{5}\left(\begin{array}{cc}
0 & 0\\
1 & 0\end{array}\right),\;\;\widehat{\Gamma}^{\dagger}=-2gi\gamma_{5}\left(\begin{array}{cc}
0 & 1\\
0 & 0\end{array}\right),\end{equation}
with $g$ the Yukawa coupling constant. We can write \[
\Phi=\varphi+\varphi_{0}\equiv\frac{1}{\sqrt{2}}(\varphi_{R}+i\varphi_{I})+\varphi_{0},\]
where $\varphi_{0}$ is the expectation value of the vacuum or the
zero mode of the boson field and $\varphi_{R,I}$ denote the real
and imaginary parts of non-zero modes. With $\varphi_{0}$ the U(1)
symmetry is spontaneously broken. To make our approach transparent
we did not include in Eq. (\ref{eq:lag01}) the self-interaction of
$\Phi$ such as $\Phi^{4}$, which does not change the results qualitatively.
The quartic or higher terms in $\Delta$ (or $\varphi_{0}$) are present
automatically from the effective potential. The Lagrangian (\ref{eq:lag01})
then becomes \begin{eqnarray}
\mathcal{L}(\Delta,\varphi,\Psi) & = & -\frac{1}{2}\overline{\Psi}S^{-1}\Psi+\frac{\mu_{b}^{2}-m_{b}^{2}}{4g^{2}}|\Delta|^{2}+|(\partial_{t}-i\mu_{b})\varphi|^{2}-|\nabla\varphi|^{2}-m_{b}^{2}|\varphi|^{2}+\frac{1}{2}\left(\varphi^{\dagger}\overline{\Psi}\widehat{\Gamma}\Psi-\varphi\overline{\Psi}\widehat{\Gamma}^{\dagger}\Psi\right)\nonumber \\
 & = & -\frac{1}{2}\overline{\Psi}S^{-1}\Psi+\frac{\mu_{b}^{2}-m_{b}^{2}}{4g^{2}}|\Delta|^{2}-\frac{1}{2}(\varphi_{R},\varphi_{I})D^{-1}\left(\begin{array}{c}
\varphi_{R}\\
\varphi_{I}\end{array}\right)+\frac{1}{2}(\varphi_{R},\varphi_{I})\overline{\Psi}\left(\begin{array}{c}
\widehat{\Gamma}_{R}\\
\widehat{\Gamma}_{I}\end{array}\right)\Psi.\label{eq:L00}\end{eqnarray}
The action is $I(\Delta,\varphi,\Psi)=\int_{X}\mathcal{L}(\Delta,\varphi,\Psi)$.
The boson-fermion vertices in the second equality are defined by \begin{eqnarray*}
\hat{\Gamma}_{R} & = & \sqrt{2}(\widehat{\Gamma}-\widehat{\Gamma}^{\dagger})=i2\sqrt{2}g\gamma_{5}\sigma_{1}^{NG},\\
\hat{\Gamma}_{I} & = & -i\sqrt{2}(\widehat{\Gamma}+\widehat{\Gamma}^{\dagger})=-i2\sqrt{2}g\gamma_{5}\sigma_{2}^{NG},\end{eqnarray*}
where $\sigma_{1,2}^{NG}$ are Pauli matrices in the Nambu-Gorkov
basis. The inverse fermion and boson propagators $S^{-1}$ and $D^{-1}$
are given in momentum space by \begin{eqnarray}
S^{-1} & = & -\left(\begin{array}{cc}
\gamma_{\mu}P^{\mu}+\mu\gamma^{0}-m & i\gamma_{5}\Delta\\
i\gamma_{5}\Delta^{*} & \gamma_{\mu}P^{\mu}-\mu\gamma^{0}-m\end{array}\right),\nonumber \\
D^{-1} & = & -\left(\begin{array}{cc}
P_{\mu}P^{\mu}+\mu_{b}^{2}-m_{b}^{2} & 2\mu_{b}ip_{0}\\
-2\mu_{b}ip_{0} & P_{\mu}P^{\mu}+\mu_{b}^{2}-m_{b}^{2}\end{array}\right),\label{eq:fermion-prop}\end{eqnarray}
where $\Delta=2g\varphi_{0}$ is the condensate. We have dropped the
mixing terms of zero and non-zero boson modes since they vanish when
carrying out the path integral. We see in (\ref{eq:L00}) that the
di-fermion boson is described by two real scalar field. 

The Lagrangian (\ref{eq:lag01}) can also be expressed in terms of
the Higgs and the Nambu-Goldstone fields by parametrizing the complex
boson field as $\Phi=\frac{1}{\sqrt{2}}(\eta+\eta_{0})e^{2i\theta}$
with $\eta+\eta_{0}=\sqrt{2}|\Phi|$ and $2\theta=\arg(\Phi)$. Here
the phase field $\theta$ is the Nambu-Goldstone field and $\eta$
the Higgs one. We can choose the unitary gauge by rewriting the fermion
field in the form and $\Psi=\left(\begin{array}{c}
e^{i\theta}\psi\\
e^{-i\theta}\psi_{C}\end{array}\right)$. Inserting the above expression for $\Phi$ and $\Psi$ into the
Lagrangian (\ref{eq:lag01}), we obtain \begin{eqnarray}
\mathcal{L}(\Delta,\varphi,\Psi) & = & -\frac{1}{2}\overline{\Psi}S^{-1}\Psi+\frac{\mu_{b}^{2}-m_{b}^{2}}{4g^{2}}\Delta^{2}+\frac{1}{2}\left[(\partial_{\nu}\eta)^{2}+(\mu_{b}^{2}-m_{b}^{2})\eta^{2}\right]\nonumber \\
 &  & +2(\eta+\eta_{0})^{2}\left[(\partial_{\nu}\theta)^{2}-\mu_{b}\partial_{0}\theta\right]-\frac{1}{2}(\partial^{\mu}\theta)\overline{\Psi}\sigma_{3}\gamma_{\mu}\Psi+\frac{1}{2}\eta\overline{\Psi}\widetilde{\Gamma}\Psi,\label{eq:L01}\end{eqnarray}
where $S^{-1}$ is given by Eq. (\ref{eq:fermion-prop}) but with
$\Delta=\sqrt{2}g\eta_{0}$ which is different from the case of (\ref{eq:L00}).
Note that in the Lagrangian (\ref{eq:L01}) we still use the same
symbol $\Psi$ and $\overline{\Psi}$ but now without the phase $\theta$:
$\Psi=\left(\begin{array}{c}
\psi\\
\psi_{C}\end{array}\right)$ and $\overline{\Psi}=(\overline{\psi},\overline{\psi}_{C})$. In
the second to last term $\sigma_{3}$ is the Pauli matrix in the NG
basis. In the last term we have defined a new vertex \[
\widetilde{\Gamma}\equiv\sqrt{2}ig\gamma_{5}\left(\begin{array}{cc}
0 & 1\\
1 & 0\end{array}\right).\]
We can see that the mass term of $\theta$ is absent in the tree level
as expected for the Nambu-Goldstone mode. 

A few remarks about the Lagrangian (\ref{eq:L00}) and (\ref{eq:L01})
are needed. We see that the complex boson field $\varphi$ in (\ref{eq:L00})
is transformed into the Higgs and Nambu-Goldstone fields in the Lagrangian
(\ref{eq:L01}). Both (\ref{eq:L00}) and (\ref{eq:L01}) are parametrizations
of the same Lagrangian (\ref{eq:lag01}) which are equivalent to each
other. We choose (\ref{eq:L00}) in this paper as it is more simple
and transparent than (\ref{eq:L01}). For example, in (\ref{eq:L00}),
in addition to the 2PI diagrams from the fermion and Higgs fields,
there are also diagrams from the fermion and Nambu-Goldstone fields
from (\ref{eq:L01}), which are much more complicated. Another example
is that the only two-point Green function for bosons from (\ref{eq:L00})
is $\left\langle P\varphi(X)\varphi^{\dagger}(Y)\right\rangle \sim\left\langle P\eta(X)e^{2i\theta(X)}\eta(Y)e^{-2i\theta(Y)}\right\rangle $,
where $P$ denotes the path-ordered operator, contains the correlations
of both the Higgs and Nambu-Goldstone fields. 

Note that we used the roman letter $S$ to denote the bare fermion
propagator but with the condensate $\Delta$, which is also called
the tree level propagator. We will use calligraphic letter $\mathcal{S}$
to denote the fully dressed fermion propagator.

\section{CJT formalism}

\begin{figure}
\caption{\label{fig:2PI-vacuum}(a) The 2PI vacuum diagrams up to 2 loops.
(b) The boson self-energy derived from the 2PI vacuum diagrams. (c)
The fermion self-energy derived from the 2PI vacuum diagrams. The
solid and dashed lines denote fermion and boson full propagators.
$X,Y$ label indices of the color, flavor, space and/or bosonic species. }

\includegraphics[scale=0.6]{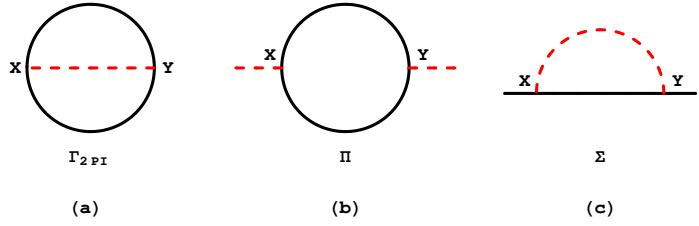}
\end{figure}

We start from the Lagrangian (\ref{eq:L00}). The CJT effective potential
reads \cite{Cornwall:1974vz}, \begin{eqnarray}
\Gamma(\Delta,\overline{\mathcal{D}},\overline{\mathcal{S}}) & = & -I(\Delta)+\frac{1}{2}\left\{ \mathrm{Tr}\ln\overline{\mathcal{D}}^{-1}+\mathrm{Tr}(D^{-1}\overline{\mathcal{D}}-1)\right.\nonumber \\
 &  & \left.-\mathrm{Tr}\ln\overline{\mathcal{S}}^{-1}-\mathrm{Tr}(S^{-1}\overline{\mathcal{S}}-1)-2\Gamma_{2PI}(\overline{\mathcal{D}},\overline{\mathcal{S}})\right\} ,\label{eq:eff-pot-cjt}\end{eqnarray}
where $\overline{\mathcal{D}},\overline{\mathcal{S}}$ are full boson
and fermion propagators and $I(\Delta)\equiv I(\Delta,0,0)$ is the
tree-level action with all fields replaced by their expectation values
(we have assumed $\left\langle \varphi\right\rangle =\left\langle \Psi\right\rangle =\left\langle \overline{\Psi}\right\rangle =0$).
The factor $1/2$ for the fermion term $-\mathrm{Tr}\ln\overline{\mathcal{S}}^{-1}$
comes from the doubling of degrees of freedom in NG basis. $D^{-1}$
and $S^{-1}$ the inverse tree-level propagators for bosons and fermions,
\begin{eqnarray}
D_{ij}^{-1} & = & -\frac{\delta^{2}I}{\delta\varphi_{i}(X)\delta\varphi_{j}(Y)},\end{eqnarray}
with $i,j=R,I$ and $S^{-1}$ is given by Eq. (\ref{eq:fermion-prop}).
Note that all propagators differ their normal forms by a sign, i.e.
$S\rightarrow-S$ and $D\rightarrow-D$ etc. The contribution of 2-particle
irreducible vacuum diagrams $\Gamma_{2}$ is truncated to the 2-loop
level as shown in Fig. \ref{fig:2PI-vacuum}, \begin{eqnarray}
\Gamma_{2PI}(\overline{\mathcal{D}},\overline{\mathcal{S}}) & \approx & -\frac{1}{4}\mathrm{Tr}\left\{ \overline{\mathcal{D}}_{XY}\mathrm{Tr}\left[\widehat{\Gamma}_{X}\overline{\mathcal{S}}_{XY}\widehat{\Gamma}_{Y}\overline{\mathcal{S}}_{YX}\right]\right\} ,\label{eq:2pi}\end{eqnarray}
where $X,Y$ denote indices of color, flavor, space, Nambu-Gorkov
and/or bosonic species, and the trace is of functional sense running
over the general indices. The full propagators are denoted by calligraphic
letters. The negative sign comes from the quark loop. A factor of
1/2 is from the Taylor expansion of the interaction and the other
one is from the NG basis. 

The Dyson-Schwinger (DS) equations for fermions and bosons can be
derived by taking derivatives of the effective potential with respect
to propagators, \begin{eqnarray}
\left.\frac{\delta\Gamma}{\delta\overline{\mathcal{D}}}\right|_{\overline{\mathcal{D}}=\mathcal{D},\overline{\mathcal{S}}=\mathcal{S}} & = & 0,\nonumber \\
\left.\frac{\delta\Gamma}{\delta\overline{\mathcal{S}}}\right|_{\overline{\mathcal{D}}=\mathcal{D},\overline{\mathcal{S}}=\mathcal{S}} & = & 0,\end{eqnarray}
which lead to the DS equations, \begin{eqnarray}
\mathcal{D}^{-1} & = & D^{-1}-2\frac{\delta\Gamma_{2PI}}{\delta\mathcal{D}}=D^{-1}+\Pi(\mathcal{D},\mathcal{S}),\nonumber \\
\mathcal{S}^{-1} & = & S^{-1}+2\frac{\delta\Gamma_{2PI}}{\delta\mathcal{S}}=S^{-1}-\Sigma(\mathcal{D},\mathcal{S}).\label{eq:dse-fermion-boson}\end{eqnarray}
where the self-energies for bosons and fermions $\Pi$ and $\Sigma$
come from $\Gamma_{2PI}$ in Eq. (\ref{eq:2pi}) and are shown in
Fig. \ref{fig:2PI-vacuum}(b) and (c) respectively. They are expressed
by, \begin{eqnarray}
\Pi(\mathcal{S},\mathcal{D}) & = & \frac{1}{2}\mathrm{Tr}[\widehat{\Gamma}\mathcal{S}\widehat{\Gamma}\mathcal{S}],\nonumber \\
\Sigma(\mathcal{D},\mathcal{S}) & = & \mathrm{Tr}[\mathcal{D}\widehat{\Gamma}\mathcal{S}\widehat{\Gamma}].\end{eqnarray}
Substituting the DS equations (\ref{eq:dse-fermion-boson}) into Eq.
(\ref{eq:eff-pot-cjt}), we arrive at \begin{eqnarray}
\Gamma(\Delta,\mathcal{D},\mathcal{S}) & = & -I(\Delta)+\frac{1}{2}\left\{ \mathrm{Tr}\ln\mathcal{D}^{-1}+\mathrm{Tr}[\mathcal{D}^{-1}\mathcal{D}-\Pi(\mathcal{D},\mathcal{S})\mathcal{D}-1]\right.\nonumber \\
 &  & \left.-\mathrm{Tr}\ln\mathcal{S}^{-1}-\mathrm{Tr}[\mathcal{S}^{-1}\mathcal{S}+\Sigma(\mathcal{D},\mathcal{S})\mathcal{S}-1]-2\Gamma_{2PI}(\mathcal{D},\mathcal{S})\right\} \nonumber \\
 & = & -I(\Delta)+\frac{1}{2}\left\{ \mathrm{Tr}\ln\left[D^{-1}(1+D\Pi)\right]-\mathrm{Tr}\ln\left[S^{-1}(1-S\Sigma)\right]-\mathrm{Tr}(\Sigma\mathcal{S})\right\} \nonumber \\
 & \approx & -I(\Delta)+\frac{1}{2}\left\{ \mathrm{Tr}\ln D^{-1}-\mathrm{Tr}\ln S^{-1}+\mathrm{Tr}[D\Pi(\mathcal{D},\mathcal{S})]\right\} .\label{eq:eff-pot-cjt1}\end{eqnarray}
Hereafter we use the thermodynamical potential $\Omega$ to replace
the effective potential $\Gamma$, $\Omega=\Omega_{0}-\Gamma_{2PI}$,
where the sum of first three terms in Eq. (\ref{eq:eff-pot-cjt1})
are denoted as $\Omega_{0}$. Note that the expansion in terms of
self-energies made in the last line somehow leads to a partial loss
of self-consistency, but it much simplifies the numerical calculation.
One can also make an expansion in terms of the condensate $\Delta$
near the critical temperature where $\Delta$ is small \cite{Giannakis:2004xt,Noronha:2006cz}.

\section{Pseudogap and renormalized boson mass}

\begin{figure}
\caption{\label{cap:pseudogap}Diagrams for the gap and pseudogap. The numbers
1 and 2 are the NG indices. }

\includegraphics[scale=0.8]{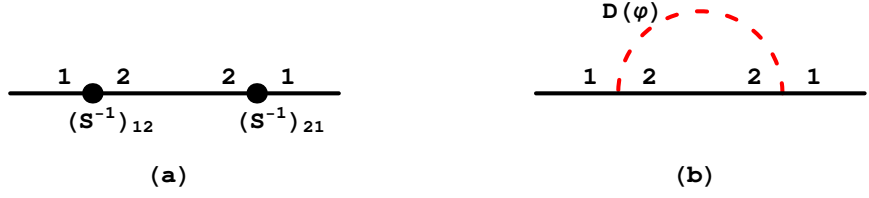}
\end{figure}

In this section we address the DS equations (\ref{eq:dse-fermion-boson})
for bosons and fermions. First we address the DS equation for fermions.
When obtaining $\mathcal{S}$ from elements of $\mathcal{S}^{-1}$,
we will encounter a term $(\mathcal{S}^{-1})_{12}[(\mathcal{S}^{-1})_{22}]^{-1}(\mathcal{S}^{-1})_{21}$
which is illustrated in Fig. \ref{cap:pseudogap}(a). We will show
that the correction to the fermion self-energy due to fluctuations
of di-fermion bosons in Fig. \ref{cap:pseudogap}(b) contributes to
the diagonal parts of the full inverse propagator $\mathcal{S}^{-1}$. 

The contribution from Fig. (\ref{cap:pseudogap}b) can be written
as \begin{eqnarray}
\Sigma_{11}(\mathcal{D},\mathcal{S}) & \approx & \left\{ \mathrm{Tr}[D_{ij}\widehat{\Gamma}_{i}S\widehat{\Gamma}_{j}]\right\} _{11}\nonumber \\
 & = & -16g^{2}T\sum_{n}\int\frac{d^{3}q}{(2\pi)^{3}}\left[D_{RR}(Q)+iD_{IR}(Q)\right]\gamma_{5}S_{22}(P+Q)\gamma_{5}\nonumber \\
 & = & 16g^{2}T\sum_{n}\int\frac{d^{3}q}{(2\pi)^{3}}\frac{1}{(p_{0}+\mu_{b})^{2}-E_{p}^{2}}\gamma_{0}\Lambda_{p+q}^{-e}\frac{1}{p_{0}+q_{0}-\mu-eE_{p+q}}\label{eq:ps01}\end{eqnarray}
where we have taken the bare propagator $S$ and $D$ in (\ref{eq:fermion-prop})
as approximations to the full ones. Here $i,j=R,I$ denote the real
and imaginary components, $q_{0}$ is the bosonic Matsubara frequency
$q_{0}=i2\pi nT$. We also uses the equalities $D_{RR}(Q)=D_{II}(Q)$
and $D_{IR}(Q)=-D_{RI}(Q)$, see App. (\ref{sec:self-en}). The energy
projector is defined by \[
\Lambda_{p}^{e}=\frac{1}{2}\left(1+e\frac{\gamma_{0}\bm{\gamma}\cdot\mathbf{p}+\gamma_{0}m}{E_{p}}\right),\]
where $e=\pm$. The presence of $\Sigma_{11/22}$ is equivalent to
$(S^{-1})_{11/22}$ gets a correction, where $S^{-1}$ is given by
Eq. (\ref{eq:fermion-prop}). We can verify it by seeing \begin{eqnarray}
\mathcal{S}_{11} & = & \left[(\mathcal{S}^{-1})_{11}-(\mathcal{S}^{-1})_{12}((\mathcal{S}^{-1})_{22})^{-1}(\mathcal{S}^{-1})_{21}\right]^{-1}\nonumber \\
 & = & \left[(S^{-1})_{11}-\Sigma_{11}-(S^{-1})_{12}((S^{-1})_{22}-\Sigma_{22})^{-1}(S^{-1})_{21}\right]^{-1}\nonumber \\
 & \approx & \left[(S^{-1})_{11}-\Sigma_{11}-(S^{-1})_{12}((S^{-1})_{22})^{-1}(S^{-1})_{21}\right]^{-1}\end{eqnarray}
where $\Sigma_{11}$ is assumed the value in Eq. (\ref{eq:ps01}).
Note that in reaching the last line of the above equation we have
neglected $\Sigma_{22}$ for clarity since it is sandwiched by $(S^{-1})_{12}$
and $(S^{-1})_{21}$ and complicates the calculation. We do not include
fermion self-energy contribution to $\Sigma_{11}$, $\sim\gamma_{0}g^{2}k_{0}\ln\frac{M^{2}}{k_{0}^{2}}$
\cite{Wang:2001aq,Schmitt:2002sc} coming from long range gauge interaction
such as magnetic photon or gluon exchanges, since the gauge field
is absent in the current version of our model. We can easily evaluate
\begin{eqnarray}
-\Sigma_{11}-(S^{-1})_{12}((S^{-1})_{22})^{-1}(S^{-1})_{21} & \approx & -16g^{2}T\sum_{n,e}\int\frac{d^{3}q}{(2\pi)^{3}}\gamma_{0}\Lambda_{p}^{-e}\frac{1}{p_{0}+q_{0}-\mu-eE_{p}}\times\frac{1}{(p_{0}+\mu_{b})^{2}-E_{p}^{2}}\nonumber \\
 &  & +\sum_{e}\frac{\Delta^{2}}{p_{0}-\mu-eE_{p}}\gamma_{0}\Lambda_{p}^{-e}\approx\sum_{e}\frac{\Delta^{2}+\Delta_{pg}^{2}}{p_{0}-\mu-eE_{p}}\gamma_{0}\Lambda_{p}^{-e},\end{eqnarray}
where we have assumed that $q\ll p$ so $|\mathbf{p}+\mathbf{q}|\approx p$.
The so-called pseudogap $\Delta_{pg}$ can be defined \cite{Kitazawa:2001ft,Kitazawa:2003cs,Pieri:2004,Chen:2005},
\begin{eqnarray}
\Delta_{pg}^{2} & = & -16g^{2}T\sum_{n}\int\frac{d^{3}q}{(2\pi)^{3}}\frac{1}{(p_{0}+\mu_{b})^{2}-(E_{p}^{b})^{2}}=16g^{2}\int\frac{d^{3}q}{(2\pi)^{3}}\frac{1+f_{B}(E_{q}^{b}-\mu_{b})+f_{B}(E_{q}^{b}+\mu_{b})}{2E_{q}^{b}}\nonumber \\
 & \sim & 16g^{2}\int\frac{d^{3}q}{(2\pi)^{3}}\frac{f_{B}(E_{q}^{b}-\mu_{b})+f_{B}(E_{q}^{b}+\mu_{b})}{2E_{q}^{b}},\label{eq:pg-def}\end{eqnarray}
where in the last line we remove the divergent part $1/(2E_{q}^{b})$
coming from vacuum. The boson energy is $E_{q}^{b}=\sqrt{q^{2}+m_{b}^{2}}$.
As lowest order contribution, the role of the DS equation for fermions
is approximately equivalent to adding a pseudogap term $\Delta_{pg}^{2}$
to the condensate square $\Delta^{2}$ in the dispersion relation
of quasi-particles. A few remarks about the pseudogap are in order.
Here we define the pseudogap as the correction to fermion selfenergy
from Fig. \ref{cap:pseudogap}(b) at the static limit. Its effect
looks like that the real gap square gets a correction, but it is not
a real gap. Actually it has analytical structure from which one can
compute the density of states from the difermion fluctuation \cite{Kitazawa:2001ft,Kitazawa:2003cs}. 

The renormalized boson mass can be determined by solving the pole
equation of the full propagator at static limit with zero gap and
pseudogap,\begin{equation}
\left.\mathrm{det}\mathcal{D}^{-1}(p_{0},p)\right|_{p=\Delta=\Delta_{pg}=0}=\mathrm{det}\left[\left.D^{-1}(p_{0},p)\right|_{p=0}+\left.\Pi(p_{0},p)\right|_{p=\Delta=\Delta_{pg}=0}\right]=0,\label{eq:boson-mass}\end{equation}
where $p_{0}$ is a real number with analytic extension. Note that
the off-diagonal parts of the boson self-energy are vanishing, $\Pi_{IR}=\Pi_{RI}=0$.
The analytical expressions for $\Pi_{RR}$ and $\Pi_{II}$ are given
in App. (\ref{sec:self-en}). When setting $\Delta=0$, we have \begin{equation}
\Pi_{0}(p_{0})\equiv\left.\Pi_{RR}(p_{0},p)\right|_{p=\Delta=0}=\left.\Pi_{II}(p_{0},p)\right|_{p=\Delta=0}.\end{equation}
Eq. (\ref{eq:boson-mass}) is then simplified as \begin{eqnarray}
\left[(p_{0}+\mu_{b})^{2}-m_{b}^{2}-\Pi_{0}(p_{0})\right]\left[(p_{0}-\mu_{b})^{2}-m_{b}^{2}-\Pi_{0}(p_{0})\right] & = & 0.\end{eqnarray}
Let us focus on the following equation \begin{equation}
(p_{0}+\mu_{b})^{2}-m_{b}^{2}-\Pi_{0}(p_{0})=0.\label{eq:pole1}\end{equation}
In general Eq. (\ref{eq:pole1}) can be understood as a complex equation
by analytic extension $p_{0}\rightarrow p_{0}-i\frac{\eta}{2}$ with
the width $\eta$, then we have \begin{eqnarray}
(p_{0}+\mu_{b})^{2}-\frac{\eta^{2}}{4}-m_{b}^{2}-\mathrm{Re}\Pi_{0}(p_{0}-i\frac{\eta}{2}) & = & 0,\nonumber \\
-(p_{0}+\mu_{b})\eta-\mathrm{Im}\Pi_{0}(p_{0}-i\frac{\eta}{2}) & = & 0.\label{eq:pole2}\end{eqnarray}
The renormalized mass for bosons can be defined by \begin{equation}
m_{br}^{2}\equiv(p_{0}+\mu_{b})^{2}=m_{b}^{2}+\mathrm{Re}\Pi_{0}(p_{0}-i\frac{\eta}{2})\end{equation}
with a small positive $\eta$. If the boson is stable, then $\eta=0$
and the above becomes \begin{equation}
m_{br}^{2}\equiv(p_{0}+\mu_{b})^{2}=m_{b}^{2}+\Pi_{0}(p_{0}).\label{eq:re-mass-boson}\end{equation}
By setting $\eta=0$, we can determine the dissociation temperature
$T^{*}$ for bosons from the first line of Eq. (\ref{eq:pole2}) with
$p_{0}=2m-\mu_{b}=2(m-\mu)>0$. It is necessary to solve the DS equation
in full consistency to get $T^{*}$, which is very involved and we
reserve it for a future study.

\section{Gap and density equation}

In this section we discuss the gap and density equations which are
derived from the thermodynamical potential $\Omega$ by taking its
derivative with respect to the gap $\Delta$ and chemical potential
$\mu$. 

The gap equation reads,\begin{equation}
\frac{\partial\Omega}{\partial\Delta}=\left\{ \frac{m_{b}^{2}-\mu_{b}^{2}}{4g^{2}}-\sum_{e=\pm}\int\frac{d^{3}k}{(2\pi)^{3}}\frac{\tanh[\epsilon_{k}^{e}/(2T)]}{2\epsilon_{k}^{e}}-\frac{\partial\Gamma_{2PI}}{\partial(\Delta^{2})}\right\} 2\Delta=0.\label{eq:gap-eq}\end{equation}
Here the excitation energy for fermions is given by \begin{equation}
\epsilon_{k}^{e}=\sqrt{(\xi_{k}^{e})^{2}+\Delta^{2}},\label{eq:en-fermion}\end{equation}
with the fermion energy $E_{k}=\sqrt{k^{2}+m^{2}}$ and $\xi_{k}^{e}=E_{k}-e\mu$.
The 2PI effective potential $\Gamma_{2PI}$ is analytically evaluated
in App. (\ref{sec:2pi-eff}). The density equation is obtained by
taking derivative of the thermodynamic potential with respect to the
fermion chemical potential, \begin{eqnarray}
n & = & -\frac{\partial\Omega}{\partial\mu}=\frac{2\mu\Delta^{2}}{g^{2}}+2\sum_{e=\pm}\int\frac{d^{3}k}{(2\pi)^{3}}\frac{e\xi_{k}^{e}}{2\epsilon_{k}^{e}}\left[f_{F}(\epsilon_{k}^{e})-f_{F}(-\epsilon_{k}^{e})\right]\nonumber \\
 &  & +2\sum_{e=\pm}\int\frac{d^{3}k}{(2\pi)^{3}}ef_{B}(E_{k}^{b}-e\mu_{b})+\frac{\partial\Gamma_{2PI}}{\partial\mu}\label{eq:density-eq01}\end{eqnarray}
We use an effective Fermi momentum $p_{F}$ to parametrize $n$ with
the relation $n=p_{F}^{3}/(3\pi^{2})$. Throughout the paper we fix
the value of the effective momentum $p_{F}=0.86$. We can define from
the total fermion number density $n$ the density fraction for fermions,
condensed and thermal bosons, and the 2PI part as \begin{eqnarray*}
\rho_{0} & \equiv & \frac{2\mu\Delta^{2}}{ng^{2}},\\
\rho_{F} & \equiv & \frac{2}{n}\sum_{e=\pm}\int\frac{d^{3}k}{(2\pi)^{3}}\frac{e\xi_{k}^{e}}{2\epsilon_{k}^{e}}\left[f_{F}(\epsilon_{k}^{e})-f_{F}(-\epsilon_{k}^{e})\right],\\
\rho_{B} & \equiv & \frac{2}{n}\sum_{e=\pm}\int\frac{d^{3}k}{(2\pi)^{3}}ef_{B}(E_{k}^{b}-e\mu_{b}),\\
\rho_{\Gamma_{2}} & \equiv & \frac{1}{n}\frac{\partial\Gamma_{2PI}}{\partial\mu},\end{eqnarray*}
which satisfies \[
\rho_{0}+\rho_{F}+\rho_{B}+\rho_{\Gamma_{2}}=1\]
We can treat $m_{b}^{2}$ as a crossover parameter and solve $\Delta$,
$\mu$ and $\Delta_{pg}$ with the gap equation (\ref{eq:gap-eq}),
the density equation (\ref{eq:density-eq01}) and the pseudogap definition
(\ref{eq:pg-def}).

\section{Numerical results}

\begin{figure}
\caption{\label{fig:first-order}The first order phase transition due to the
di-fermion fluctuations.}

\includegraphics[scale=0.5]{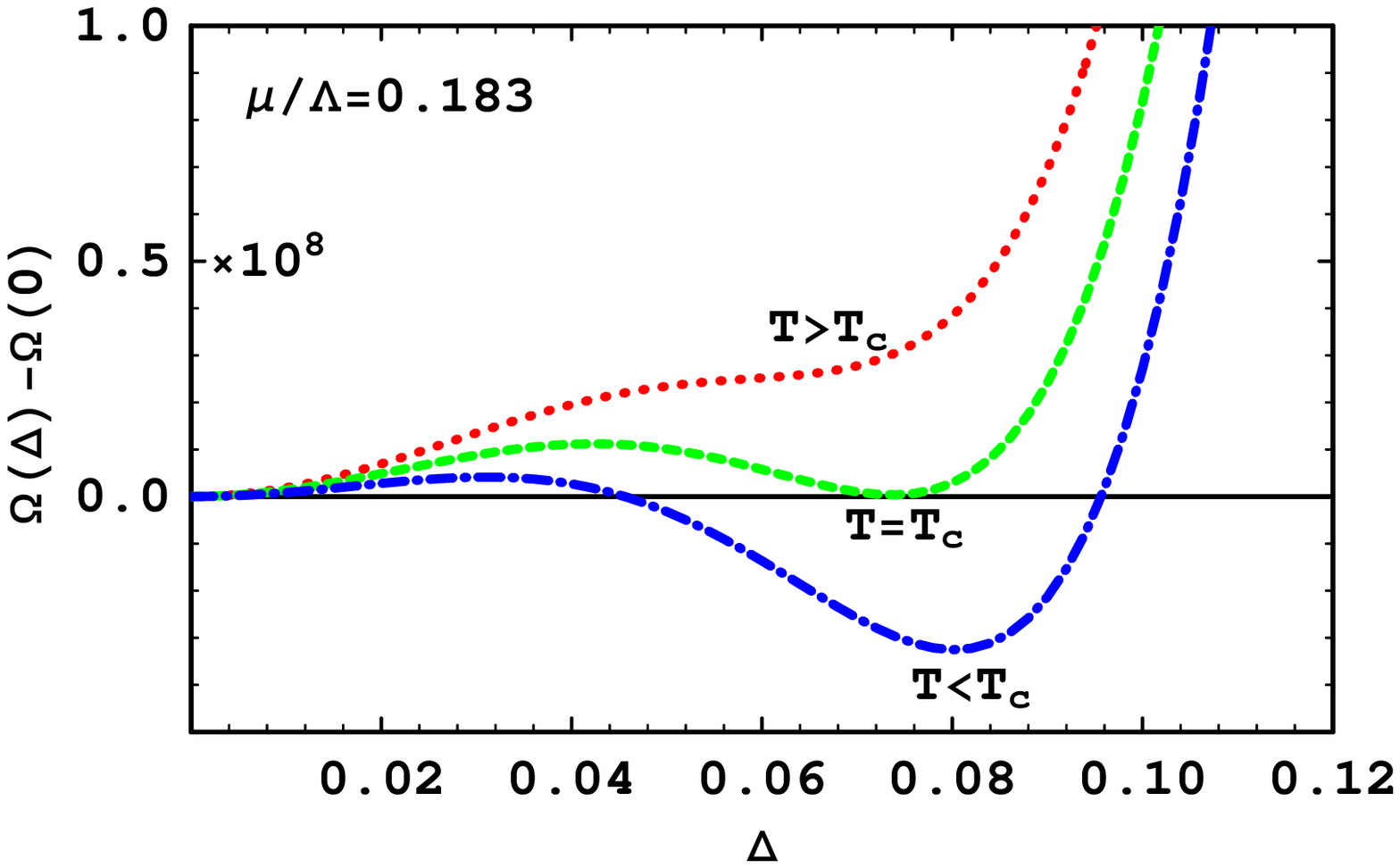}\includegraphics[scale=0.5]{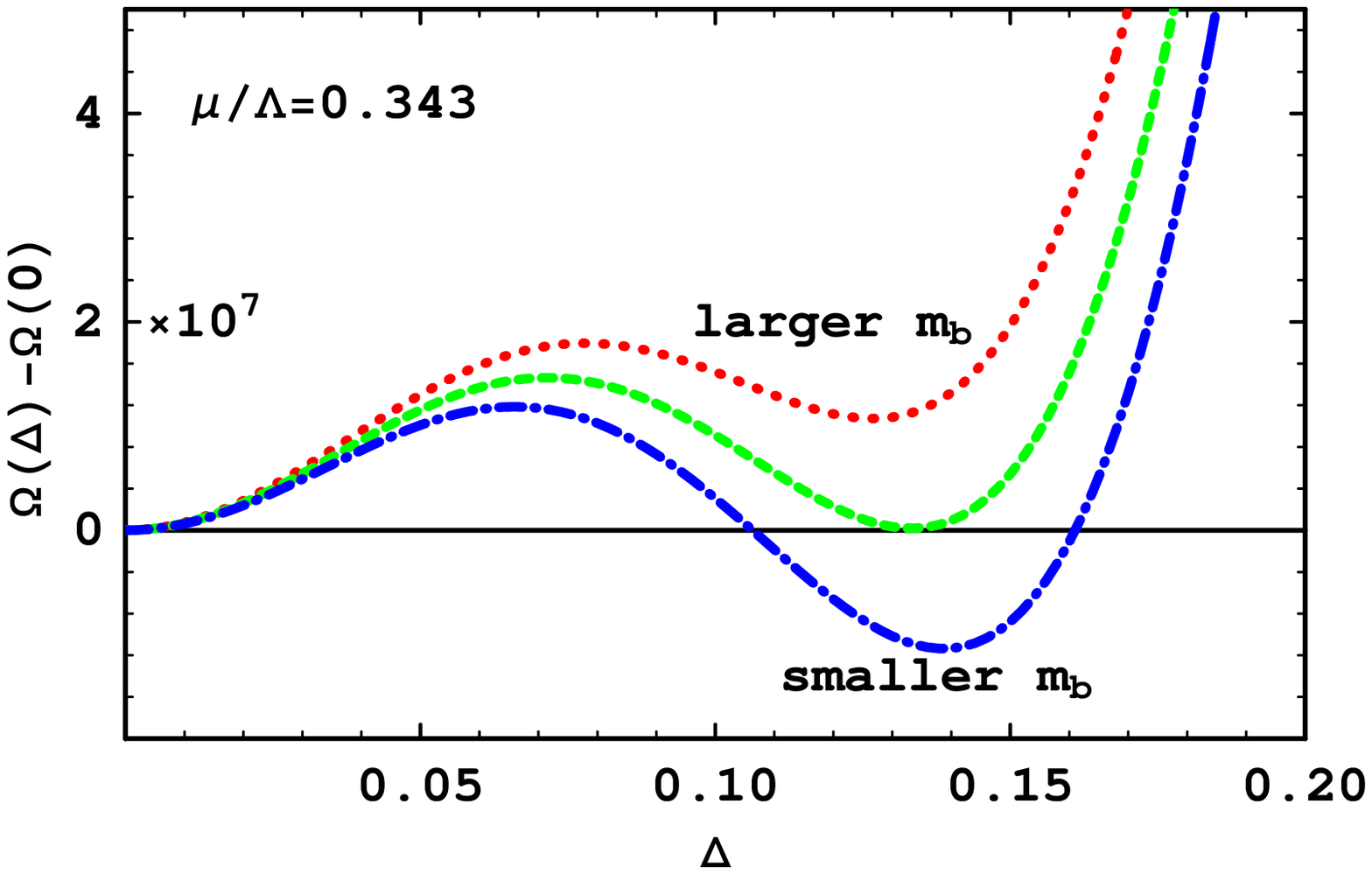}
\end{figure}

It is well known that fluctuations may change the superconducting
phase transition to be of first-order, such as the intrinsic fluctuating
magnetic field in normal superconductors \cite{Halperin:1973jh},
or the gauge field fluctuations in color superconductors \cite{Giannakis:2004xt},
due to the fact that fluctuations bring a cubic term of the condensate
to the effective potential making the Landau theroy of continuous
phase transition invalid. In our case the fluctuations of di-fermions
in superfluids also leads to a first-order phase transition. In grand
canonical ensemble with chemical potential fixed, as shown in Fig.
\ref{fig:first-order}, the thermodynamic potential $\Omega(\Delta)$
as a function of the condensate $\Delta$, has a metastable state
if the fluctuation contribution is taken into account. In the left
penal with decreasing $T$, in the right penal with decreasing $m_{b}$,
we see a sudden jump of the minimum of $\Omega(\Delta)$ from $\Delta=0$
to a non-vanishing value of $\Delta$, signaling a phase transition
from normal phase to a superfluid one. The discontinuity in the order
parameter marks the first-order phase transition.

We find that there are three extrema in $\Omega(\Delta)$ near the
transition point, which means that there are at least three solutions
to the gap equation with fixed chemical potential. So satisfying the
gap equation is not sufficient for the ground state which corresponds
to the global minimum of the thermodynamic potential.

From the density equation (\ref{eq:density-eq01}), the discontinuity
in the gap leads to that of the particle number density. This situation
cannot happen in canonical ensemble with fixed particle number. We
have checked with the free energy $F=\Omega+\mu n$ as a function
of the condensate and found no metastable state. The free energy has
only one extremum, which is the global minimum and changes continuously
at the transition point. The non-zero solution to the gap equation
corresponds to the superfluid phase, the ground state of the system.
The Thouless criterion can be employed to determine the transition
temperature. Although the order parameter changes smoothly, the phase
transition is still a first-order one. We can take the formation of
BEC in a free bosonic gas as an example. Below the critical temperature,
the chemical potential equals to the ground state energy (zero in
non-relativistic or the boson mass in relativistic system), so the
entropy per particle $s_{1}=-\left.\frac{\partial\mu}{\partial T}\right|_{P}(T<T_{c})=0$.
Above the critical temperature, the chemical potential drops with
a negative slope $\left.\frac{\partial\mu}{\partial T}\right|_{P}(T>T_{c})<0$
giving a non-zero entropy per particle $s_{2}$. So there is a nonzero
latent heat $T_{c}(s_{2}-s_{1})$. The same phenomenon occurs to our
current case with fixed total particle number. 

Now we present the numerical results by solving the gap equation (\ref{eq:gap-eq}),
the density equation (\ref{eq:density-eq01}) and the pseudogap equation
(\ref{eq:pg-def}) with fixed total particle number. From the solutions
we can also determine the transition temperature $T_{c}$ at which
the condensate $\Delta$ turns to zero. 

\begin{figure}
\caption{The gap $\Delta$, the fermion chemical potential $\mu$, the pseudogap
$\Delta_{pg}$ (the left panel) and density fractions for BEC $\rho_{0}$,
thermal bosons $\rho_{B}$, fermions $\rho_{F}$, and 2PI part $\rho_{\Gamma_{2}}$
(the right panel) as functions of $m_{b}$ with the fixed coupling
constant $g=1.8$, the temperature $T=0.14$ and the fermion mass
$m=0.28$. \label{fig:gap-den}}

\includegraphics[scale=0.5]{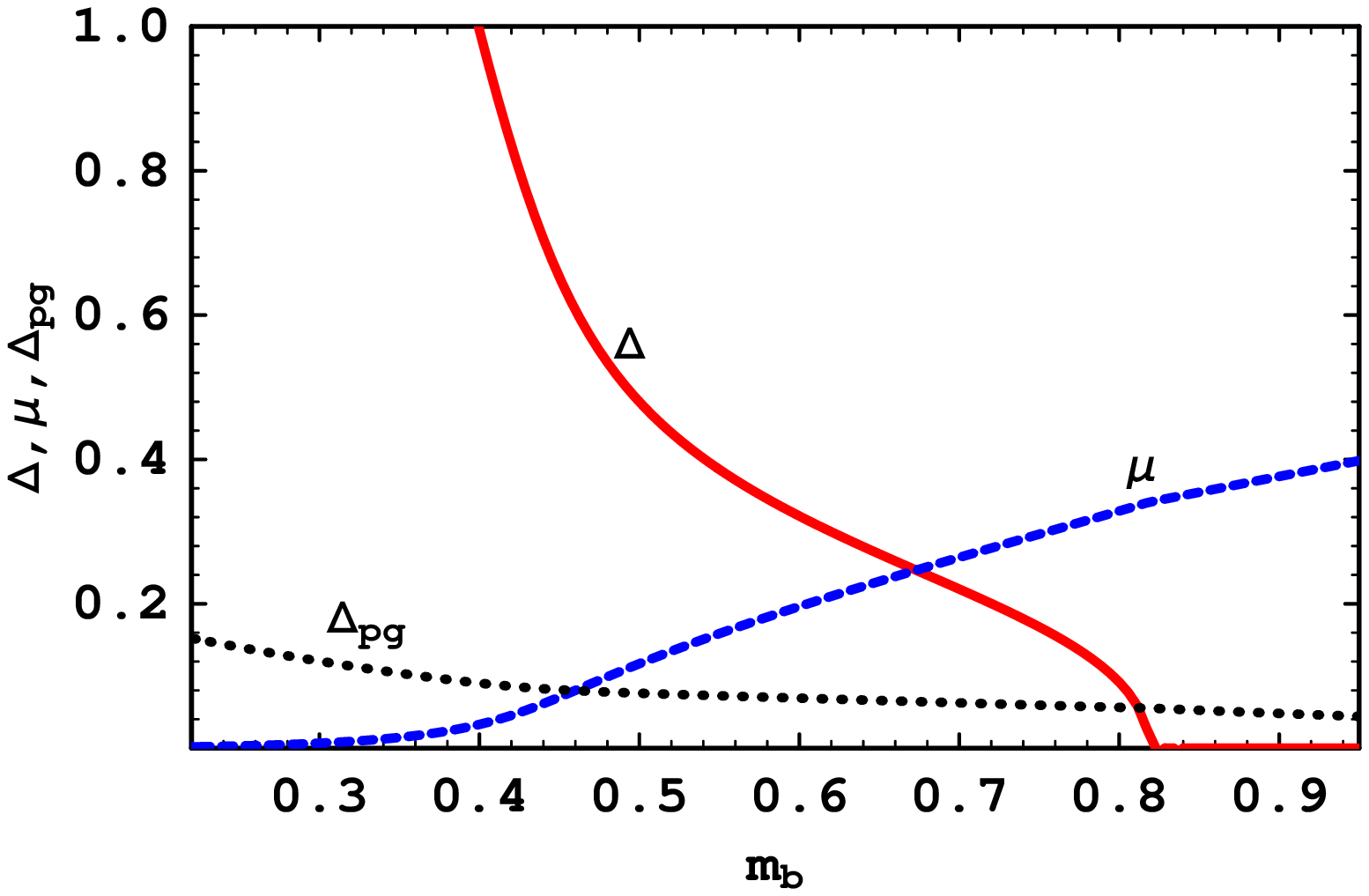}\includegraphics[scale=0.5]{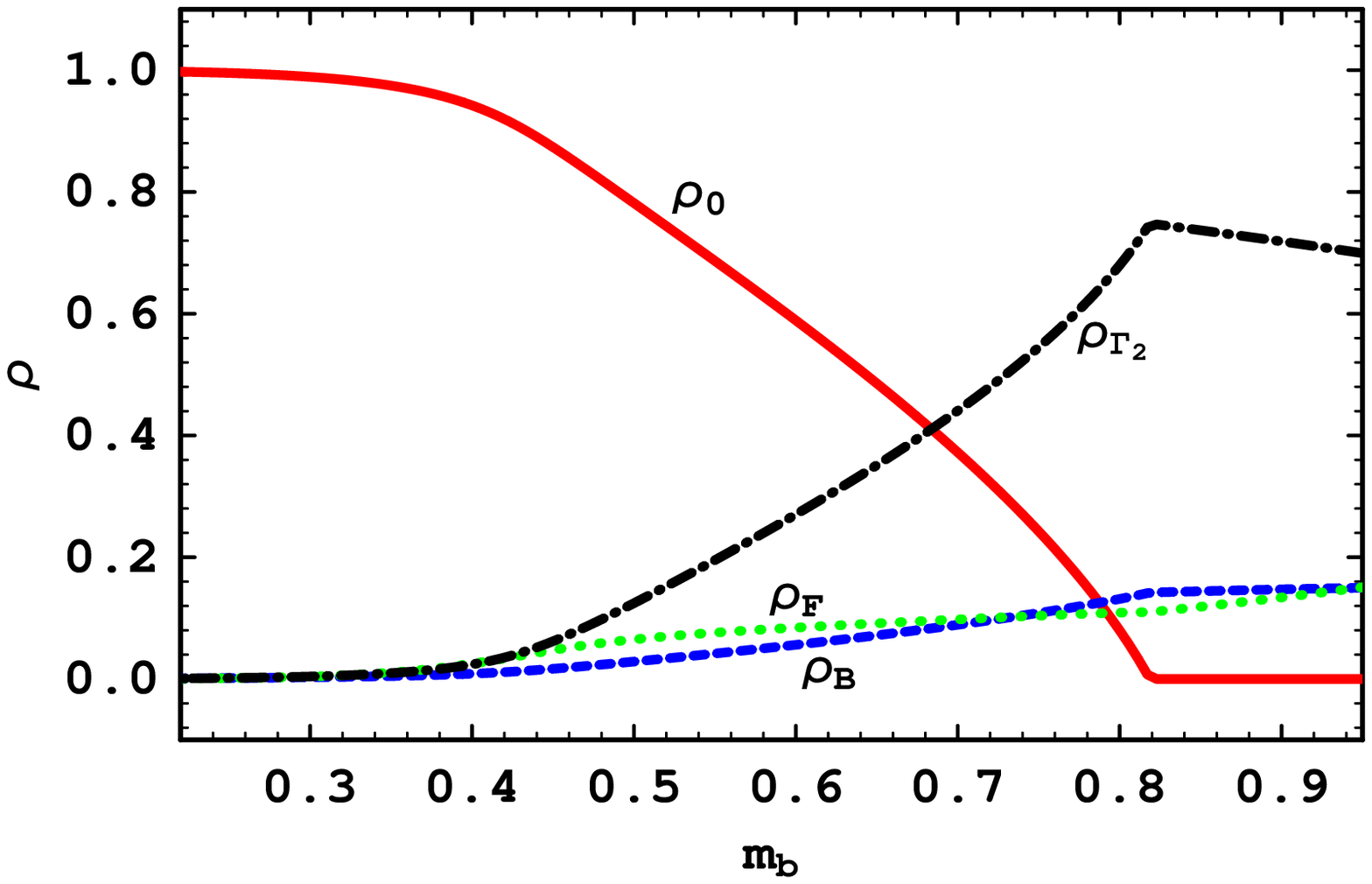}
\end{figure}

The results for the gap $\Delta$, the fermion chemical potential
$\mu$, the pseudogap $\Delta_{pg}$ and the density $\rho$ as functions
of the crossover parameter $m_{b}$ are shown in Fig. \ref{fig:gap-den}.
The parameters are set to $g=1.8$, $T=0.14$ and $m=0.28$. We see
that the BEC/BCS regions correspond to the small/large end of $m_{b}$.
In the BEC region when $m_{b}$ is small, the particle population
is dominated by the BEC part $\rho_{0}$ while the fermion part $\rho_{F}$
is negligible, and there is no Fermi surface characterized by the
small fermion chemical potential $\mu$. In the BCS region when $m_{b}$
is large, the system is mainly fermionic manifested by a clear Fermi
surface and large fraction of fermion population. We see that the
pseudogap $\Delta_{pg}$ is small due to the small temperature chosen.
The thermal density fractions $\rho_{B}$ and $\rho_{F}$ increases
with the growing chemical potential. The contribution from the 2PI
part $\rho_{\Gamma_{2}}$ is large indicating the strong interaction
between fermions and bosons. In the figure, there are breakpoints
at $m_{b}\approx0.82$ for all curves because the system enters the
normal state, i.e. $\Delta=0$, when $m_{b}$ is larger. In other
words, the breakpoints at $m_{b}\approx0.82$ indicates that the transition
temperature $T_{c}$ equals to the temperature parameter $T=0.14$.
When $m_{b}\lesssim0.82$, the left area to the break point, we have
$T_{c}\geq T=0.14$. 

\begin{figure}
\caption{Left panel: the transition temperature $T_{c}$, the fermion chemical
potential $\mu$ at $T=T_{c}$, and the pseudogap $\Delta_{pg}$ at
$T=T_{c}$ versus $m_{b}$. Right panel: density fractions for thermal
bosons $\rho_{B}$, fermions $\rho_{F}$, and the 2PI part $\rho_{\Gamma_{2}}$
all at $T=T_{c}$ versus $m_{b}$. The parameters are $g=1.8$ and
$m=0.28$. \label{fig:tran-temp}}

\includegraphics[scale=0.5]{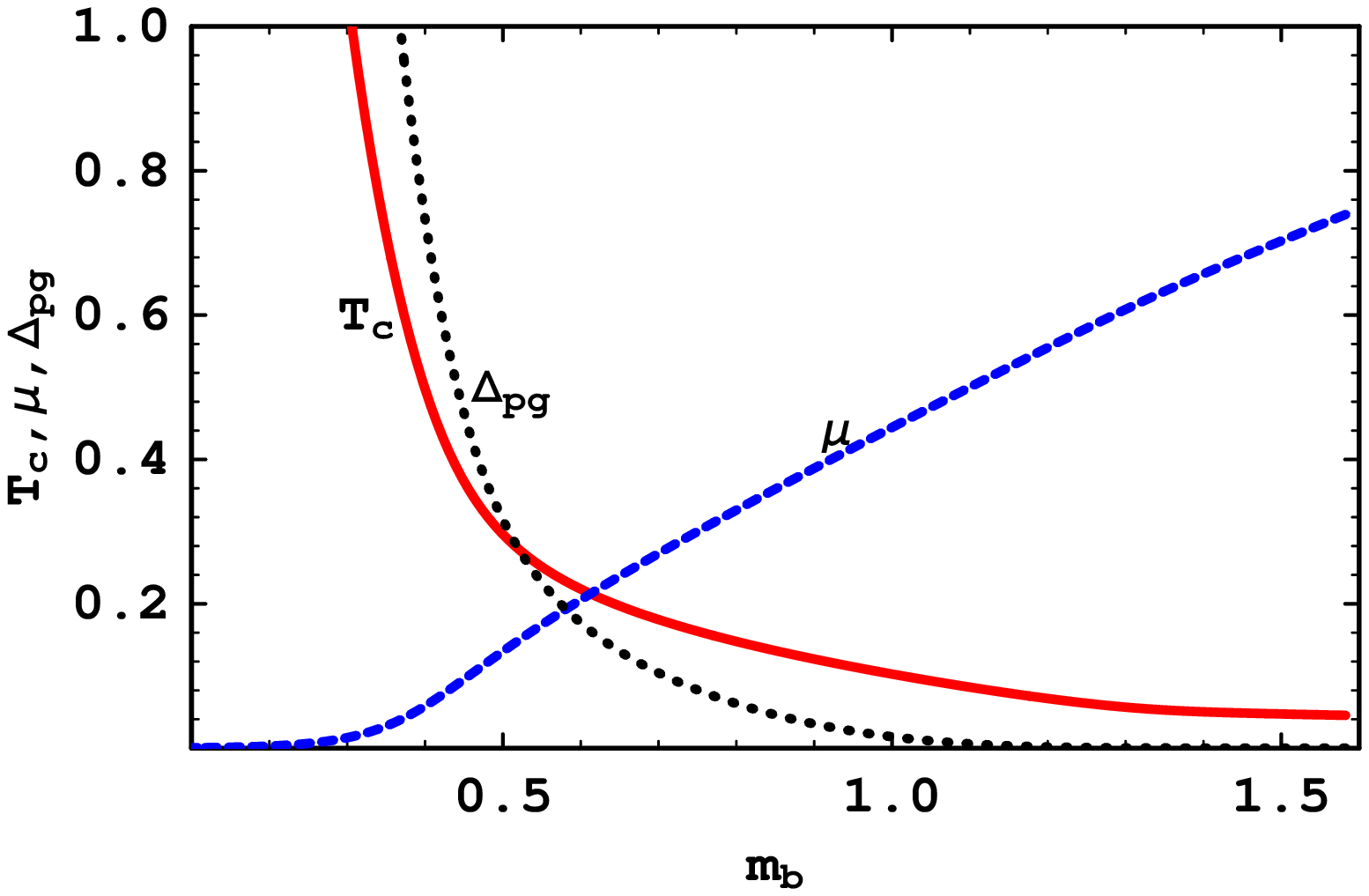}\includegraphics[scale=0.5]{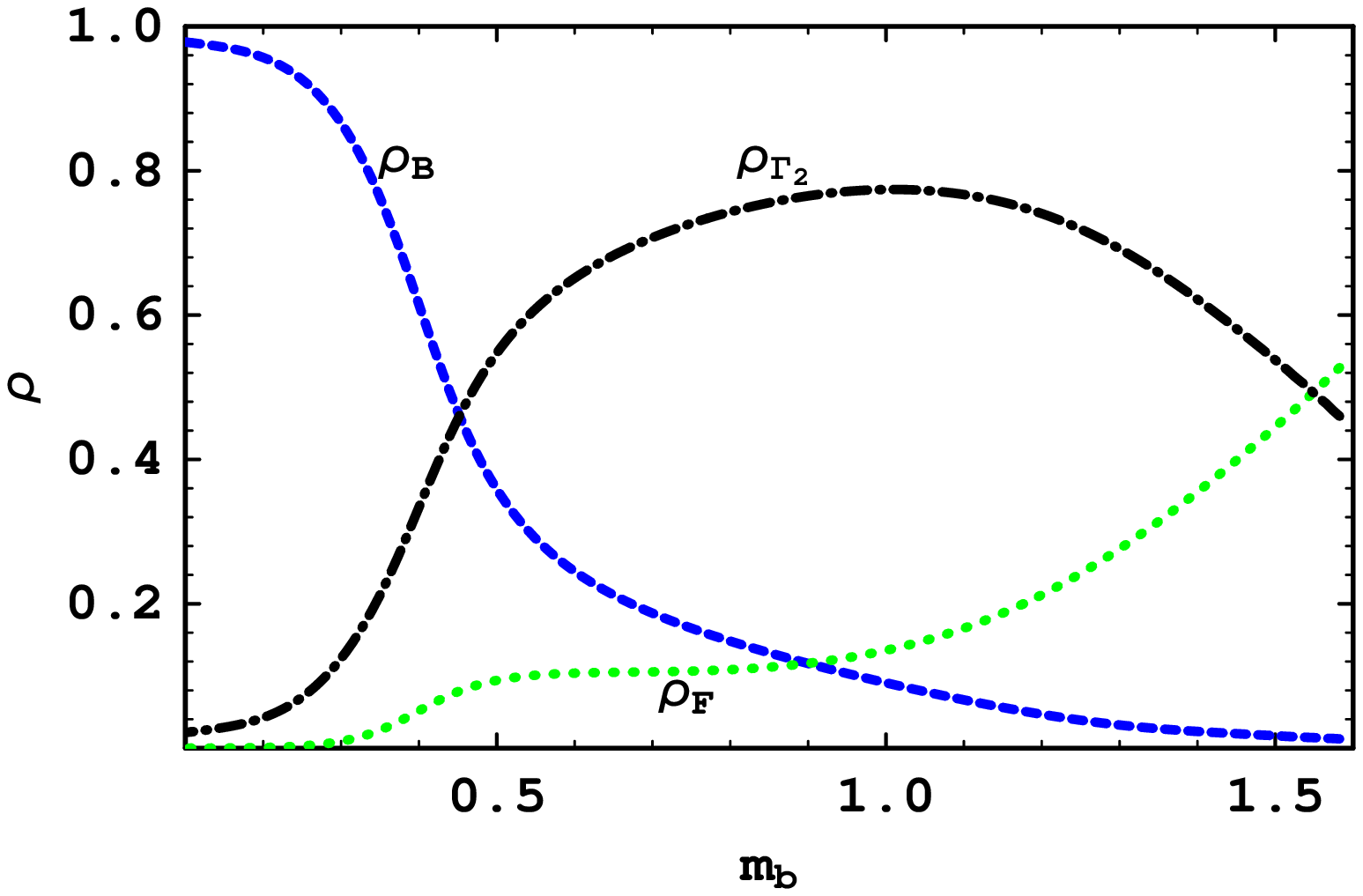}
\end{figure}

In the left panel of Fig. \ref{fig:tran-temp}, we present the results
for $T_{c}$, $\mu$ (at $T=T_{c}$) and $\Delta_{pg}$ (at $T=T_{c}$)
as functions of $m_{b}$; in the right panel we give the results for
density fractions $\rho_{B}$, $\rho_{F}$, and $\rho_{\Gamma_{2}}$
all at $T=T_{c}$ versus $m_{b}$. The behavior of $T_{c}$ is similar
to $\Delta$, comparing Fig. \ref{fig:gap-den} and Fig. \ref{fig:tran-temp}.
Although the shape of $\mu$ at $T=T_{c}$ is like $\mu$ at $T=0.14$
in Fig. \ref{fig:gap-den}, note that the temperature at each point
of the curve of $\mu$ at $T=T_{c}$ versus $m_{b}$ is different
because $T_{c}$ itself is running with $m_{b}$. Comparing to the
result for $\Delta_{pg}$ in Fig. \ref{fig:gap-den}, the result at
$T=T_{c}$ is rather large especially in the BEC region where $T_{c}$
is large. The population fraction for thermal bosons $\rho_{B}$ at
$T_{c}$ changes into $\rho_{\Gamma_{2}}$ and $\rho_{F}$ with increasing
$m_{b}$. $\rho_{B}$ intersects with $\rho_{\Gamma_{2}}$ at $m_{b}\approx0.4$
meaning a transition from a bosonic system to a strongly interacting
one. It intersects with $\rho_{F}$ at $m_{b}\approx0.9$ meaning
a transition to a fermionic interacting system. The intersection of
$\rho_{\Gamma_{2}}$ and $\rho_{F}$ occurs at $m_{b}\approx1.5$
indicating a transition from a fermionic interacting system to a fermionic
system.  

\begin{figure}
\caption{Left panel: the condensate $\Delta$, the fermion chemical potential
$\mu$, and the pseudogap $\Delta_{pg}$ versus $T$. Right panel:
density fractions for condensed and thermal bosons $\rho_{0}$ and
$\rho_{B}$, fermions $\rho_{F}$, and the 2PI part $\rho_{\Gamma_{2}}$
versus $T$. The parameters are set to $g=1.8$ and $m=0.28$ and
$m_{b}=0.56$. \label{fig:t-curve}}

\includegraphics[scale=0.5]{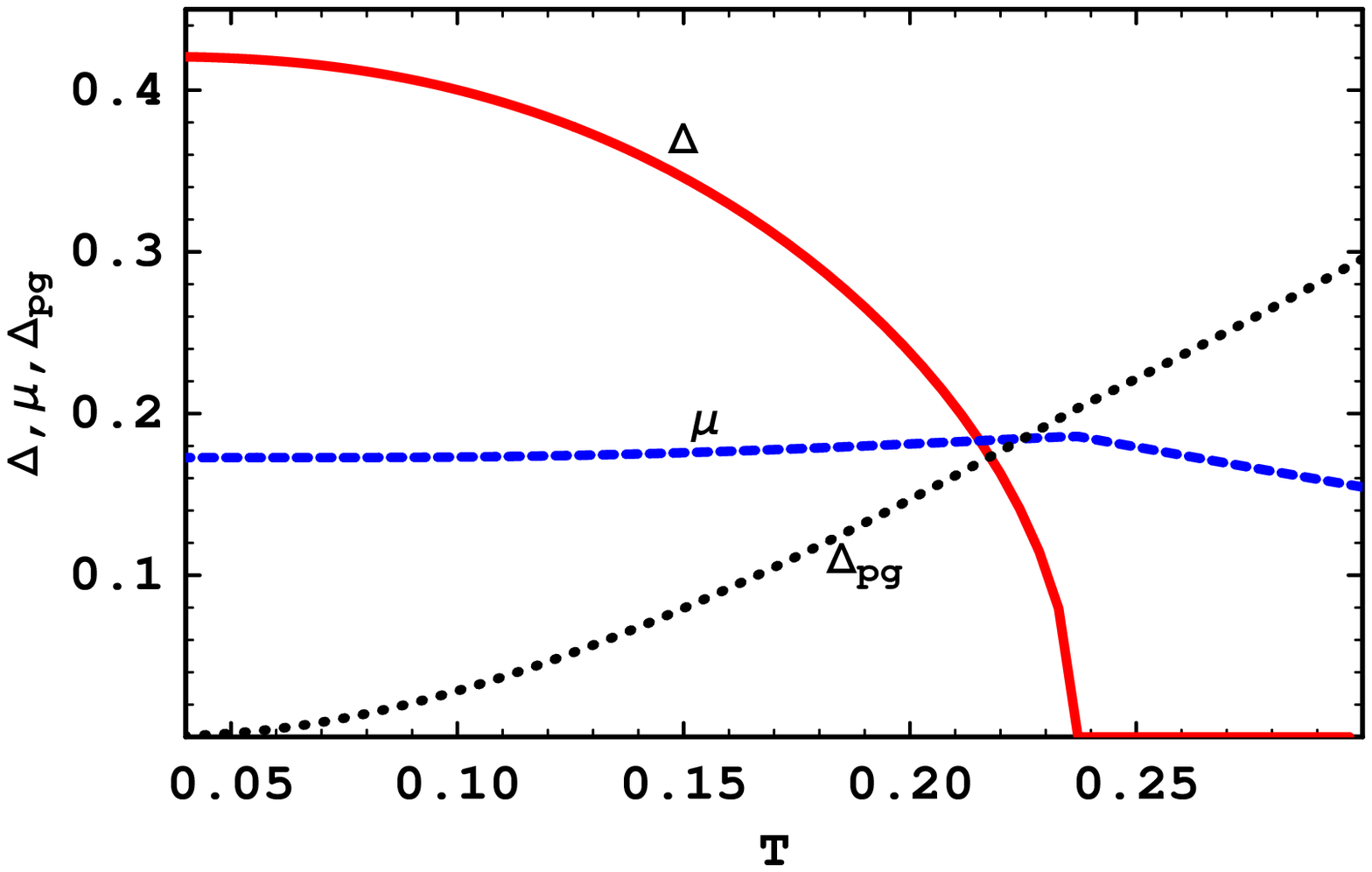}\includegraphics[scale=0.5]{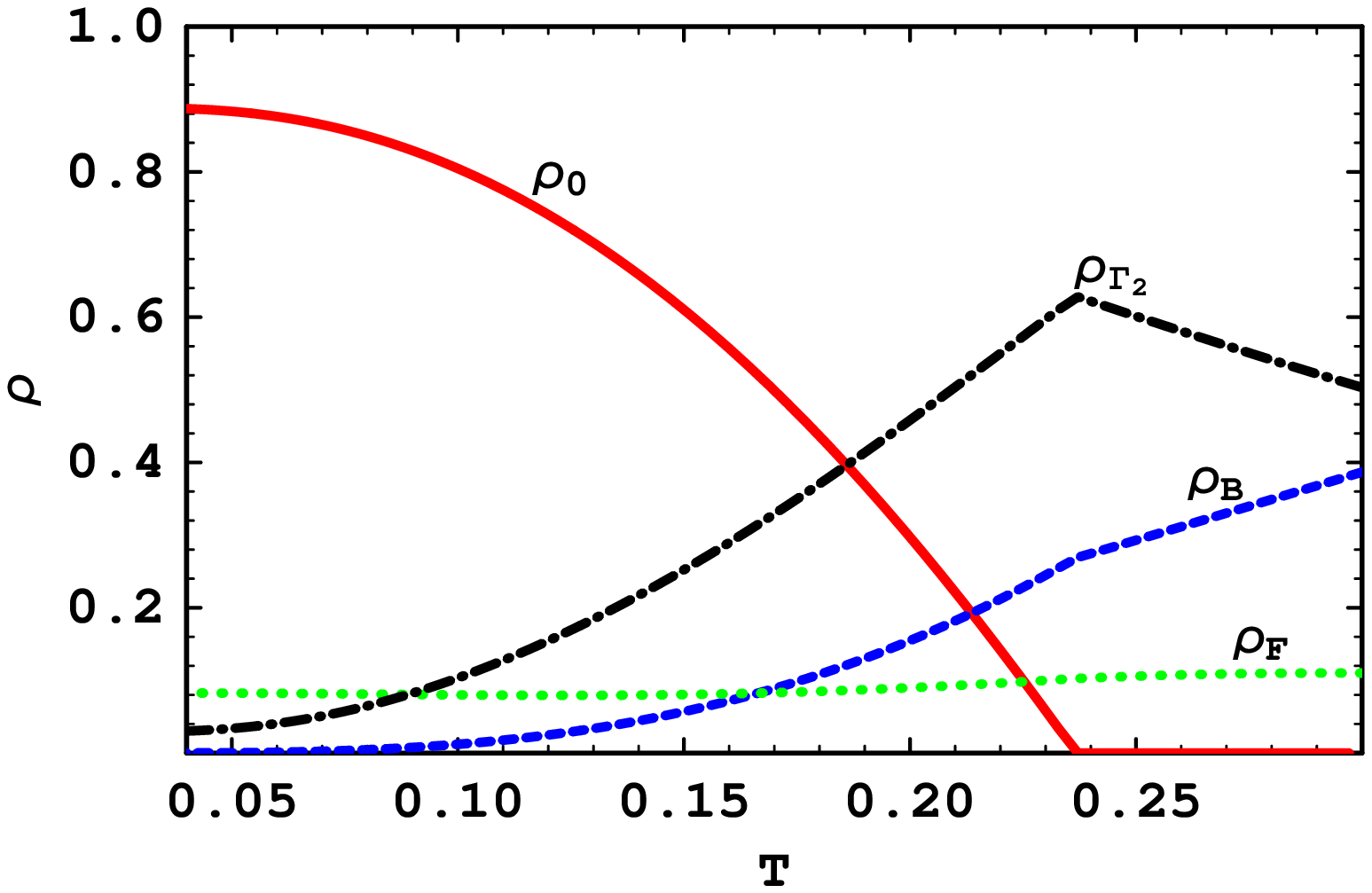}
\end{figure}

The temperature behavior of  all quantities in Fig. \ref{fig:gap-den}
and \ref{fig:tran-temp} are given in Fig. \ref{fig:t-curve} with
fixed $m_{b}=0.56$ (in BEC regime). It is easy to understand the
curve of $\Delta$ versus $T$: above a critical temperature $T_{c}$
the condensate turns to zero. There is a break point on the curve
of $\mu$ at $T_{c}$, below which $\mu$ increases with increasing
$T$ but above which it decreases. The pseudogap is a rising function
of $T$. The density fraction $\rho_{0}$ is proportional to $\Delta^{2}$,
so its shape is close to that of $\Delta$. As $T$ increases more
and more condensed bosons turn to thermal bosons, fermions and interaction
part. The thermal boson fraction $\rho_{B}$ always grows with $T$.
The 2PI part $\rho_{\Gamma_{2}}$ is also shown whose effect is large
at high $T$. Due to more sensitive response to increasing temperature
for bosons than fermions, the rising feature of $\rho_{B}$ above
$T_{c}$ can be easily understood. The decreasing of $\mu_{b}$ above
$T_{c}$ is due to particle number conservation. 

\section{Summary and discussions}

We investigate the fluctuation effects of di-fermion fields in the
BCS-BEC crossover in a relativistic superfluid within the CJT formalism.
We found in grand canonical ensemble the fluctuations lead to the
normal-superfluid phase transition of first order with metastable
states. In canonical ensemble, the phase transition is a first order
one too but without metastable states. The order parameter changes
smoothly near the transition point which ensures the validity of the
gap equation and the Thouless criterion. 

In this work, we use the bare boson mass squared $m_{b}$ as the crossover
parameter. We find that the system is in the BCS/BEC regime for large/small
$m_{b}$. Comparing to the results in our previous work \cite{Deng:2006ed}
with the crossover parameter $x=-\frac{m_{br}^{2}-\mu_{b}^{2}}{4g^{2}}$
where negative/positive $x$ defines the BCS/BEC regime, we find that
these two parameters are equivalent in describing the crossover behavior.
We can verify numerically that there is a one-to-one mapping between
$m_{b}$ and $x$. In this paper, we choose $m_{b}$ as it is more
simple. The gap equation is derived with fixed $m_{b}$ instead of
$x$ which is regarded as a derivative parameter. 

We see that the pseudo-gap $\Delta_{pg}$ and the 2PI contribution
to particle density $\rho_{\Gamma_{2}}$ vanish as the temperature
goes to zero. Therefore the results at low temperatures reproduce
those in the mean field approximation. This is because as the temperature
goes to zero the interaction between thermal bosons and fermions is
switched off which results in vanishing fluctuation contributions.
At finite temperature and in the middle of the crossover, $\rho_{\Gamma_{2}}$
is dominant indicating the strong coupling between bosons and fermions.
In the BEC or BCS end the number density is dominated by bosons or
fermions respectively, during the crossover from BCS to BEC, the system
undergoes a strongly interacting intermediate stage. This picture
is a little different from that in the mean field approach where bosons
and fermions are transformed to each other directly. 

Since the current model is a relativistic one, the anti-particles
should appear in some circumstances. In the left panel of Fig. \ref{fig:gap-den},
we see that $\mu=T=0.14$ at $m_{b}\approx0.5$, which means the thermal
energy is comparable to the energy needed for exciting a pair of fermion
and anti-fermion. At the deep end of BEC region the fraction of anti-particles
is not negligible, which is partially responsible for the large condensate.
In the left panel of Fig. \ref{fig:tran-temp}, at $m_{b}\approx0.6$
where $T_{c}=\mu$, the appearance of anti-particles drives the critical
temperature and the pseudo-gap to be large. As expected the fluctuation
effects grow with increasing temperatures. The pseudo-gap is a monotonous
increasing function of the temperature, whose magnitude is seeable
at critical temperature in the BEC regime.

In this work, we expressed the effective potential in terms of bare
propagators. The role of the DS equations are taken by the pseudo-gap
and the renormalized boson mass. There is another approach of using
the full propagators to construct the effective potential, where the
complete form of the renormalization can be implemented. Another hard
but interesting problem to solve the DS equations with full self-consistency.
All these problems deserve a detailed study in the future. 

\begin{acknowledgments} We thank D. Blaschke, M. Kitazawa, H.-c. Ren, D. Rischke, A. Schmitt and I. Shovkovy for critically reading the manuscript and for insightful discussions. Q.W. is supported in part by '100 talents' project of Chinese Academy of Sciences (CAS), by National Natural Science Foundation of China (NSFC) under the grants 10675109 and 10735040, and by the national Laboratory for the heavy Ion Accelerator in Lanzhou (NLHIAL) under the CSRm R\&D sub-project. 
\end{acknowledgments} 

\appendix

\section{Boson propagator and self-energy}

\label{sec:self-en}The bare propagator for bosons can be found from
Eq. (\ref{eq:fermion-prop}), \begin{eqnarray}
D & = & -\frac{1}{\det(D^{-1})}\left(\begin{array}{cc}
P_{\mu}P^{\mu}+\mu_{b}^{2}-m_{b}^{2} & -2\mu_{b}ip_{0}\\
2\mu_{b}ip_{0} & P_{\mu}P^{\mu}+\mu_{b}^{2}-m_{b}^{2}\end{array}\right),\end{eqnarray}
where \begin{equation}
\det(D^{-1})=[(p_{0}+\mu_{b})^{2}-(E_{p}^{b})^{2}][(p_{0}-\mu_{b})^{2}-(E_{p}^{b})^{2}]=[p_{0}^{2}-(\mu-E_{p}^{b})^{2}][p_{0}^{2}-(\mu+E_{p}^{b})^{2}].\end{equation}
We see that $D_{RR}(Q)=D_{II}(Q)$ and $D_{IR}(Q)=-D_{RI}(Q)$.

The boson self-energy matrix \begin{eqnarray}
\Pi_{ij}(\mathcal{S},\mathcal{D}) & \approx & \frac{1}{2}\mathrm{Tr}[\widehat{\Gamma}_{i}S\widehat{\Gamma}_{j}S]\end{eqnarray}
are evaluated as \begin{eqnarray}
\Pi_{RR} & \approx & \frac{1}{2}\mathrm{Tr}[\widehat{\Gamma}_{R}S\widehat{\Gamma}_{R}S]=-8g^{2}\left[\mathrm{Tr}(\gamma_{5}S_{11}\gamma_{5}S_{22})+\mathrm{Tr}(\gamma_{5}S_{21}\gamma_{5}S_{21})\right],\nonumber \\
\Pi_{II} & \approx & \frac{1}{2}\mathrm{Tr}[\widehat{\Gamma}_{I}S\widehat{\Gamma}_{I}S]=-8g^{2}\left[\mathrm{Tr}(\gamma_{5}S_{11}\gamma_{5}S_{22})-\mathrm{Tr}(\gamma_{5}S_{21}\gamma_{5}S_{21})\right],\nonumber \\
\Pi_{RI} & = & \Pi_{IR}=0,\label{eq:self-en-rr-ii}\end{eqnarray}
where \begin{eqnarray}
\mathrm{Tr}[\gamma_{5}S_{11}\gamma_{5}S_{22}] & = & -T\sum_{n}\int_{q}\frac{p_{0}-\mu+eE_{p}}{p_{0}^{2}-(\mu-eE_{p})^{2}-\Delta^{2}}\times\frac{p_{0}+q_{0}+\mu+e'E_{p+q}}{(p_{0}+q_{0})^{2}-(\mu+e'E_{p+q})^{2}-\Delta^{2}}\mathrm{Tr}[\Lambda_{p}^{e}\Lambda_{p+q}^{-e'}],\nonumber \\
\mathrm{Tr}[\gamma_{5}S_{21}\gamma_{5}S_{21}] & = & \mathrm{Tr}[\gamma_{5}S_{12}\gamma_{5}S_{12}]\nonumber \\
 & = & -T\sum_{n}\int_{q}\frac{1}{q_{0}^{2}-(\mu-eE_{q})^{2}-\Delta^{2}}\times\frac{1}{(p_{0}+q_{0})^{2}-(\mu-e'E_{p+q})^{2}-\Delta^{2}}\Delta^{2}\mathrm{Tr}[\Lambda_{q}^{-e}\Lambda_{p+q}^{-e'}].\end{eqnarray}

The results for $\Pi(p_{0},p)$ are\begin{eqnarray*}
\Pi_{RR}(p_{0},p) & = & \Pi_{0}+\Pi_{1}\\
\Pi_{II}(p_{0},p) & = & \Pi_{0}-\Pi_{1}\end{eqnarray*}
where \begin{eqnarray}
\Pi_{0}(p_{0},p) & = & -8g^{2}\sum_{e,e',e_{1},e'_{1}}\int\frac{d^{3}q}{(2\pi)^{3}}\mathrm{Tr}[\Lambda_{q}^{e}\Lambda_{p+q}^{e'}]\frac{\exp\left[\beta(-p_{0}+e'_{1}\epsilon_{p+q}^{e'}-e_{1}\epsilon_{q}^{e})\right]-1}{-p_{0}+e'_{1}\epsilon_{p+q}^{e'}-e_{1}\epsilon_{q}^{e}}\nonumber \\
 &  & \times f_{F}(-e_{1}\epsilon_{q}^{e})f_{F}(e'_{1}\epsilon_{p+q}^{e'})\frac{(\epsilon_{p+q}^{e'}-e'_{1}\mu+e'e'_{1}E_{p+q})(\epsilon_{q}^{e}+e_{1}\mu-ee_{1}E_{q})}{4\epsilon_{p+q}^{e'}\epsilon_{q}^{e}}\nonumber \\
\Pi_{1}(p_{0},p) & = & -8g^{2}\Delta^{2}\sum_{e,e',e_{1},e'_{1}}e_{1}e'_{1}\int\frac{d^{3}q}{(2\pi)^{3}}\mathrm{Tr}[\Lambda_{q}^{e}\Lambda_{p+q}^{e'}]\frac{\exp\left[\beta(-p_{0}+e'_{1}\epsilon_{p+q}^{e'}-e_{1}\epsilon_{q}^{e})\right]-1}{-p_{0}+e'_{1}\epsilon_{p+q}^{e'}-e_{1}\epsilon_{q}^{e}}\nonumber \\
 &  & \times f_{F}(-e_{1}\epsilon_{q}^{e})f_{F}(e'_{1}\epsilon_{p+q}^{e'})\frac{1}{4\epsilon_{p+q}^{e'}\epsilon_{q}^{e}}\label{eq:boson-self-en}\end{eqnarray}
where the quasi-particle energy for fermions is given by Eq. (\ref{eq:en-fermion}).
Note that in renormalized quantities/constants and the effective potential
$\Gamma_{2PI}$ the full fermion propagators are used, which amounts
to taking the pseudogap contribution into account, thus we have \begin{equation}
\epsilon_{k}^{e}\rightarrow\sqrt{(\xi_{k}^{e})^{2}+\Delta^{2}+\Delta_{pg}^{2}}.\end{equation}
In this and the next appendix, we always imply the above replacement.

\section{2PI effective potential}

\label{sec:2pi-eff}In this appendix we give the 2PI effective potential
$\Gamma_{2PI}$ using the boson self-energy in Eq. (\ref{eq:self-en-rr-ii})
as follows, \begin{eqnarray}
\Gamma_{2PI} & = & -\frac{1}{2}\mathrm{Tr}(D_{RR}\Pi_{RR})-\frac{1}{2}\mathrm{Tr}(D_{II}\Pi_{II})=-\mathrm{Tr}(D_{RR}\Pi_{0})\nonumber \\
 & = & -T\sum_{n}\int\frac{d^{3}p}{(2\pi)^{3}}\int_{0}^{\beta}d\tau'D_{RR}(\tau',p)e^{p_{0}\tau'}\int_{0}^{\beta}d\tau\Pi_{0}(\tau,p)e^{p_{0}\tau}\nonumber \\
 & = & -\int\frac{d^{3}p}{(2\pi)^{3}}\int_{0}^{\beta}d\tau D_{RR}(\beta-\tau,p)\Pi_{0}(\tau,p)\nonumber \\
 & = & 4g^{2}\int\frac{d^{3}p}{(2\pi)^{3}}\int\frac{d^{3}q}{(2\pi)^{3}}\sum_{e,e',e_{1},e'_{1},e_{2},e_{3}}\mathrm{Tr}[\Lambda_{q}^{e}\Lambda_{p+q}^{e'}]\frac{\exp[\beta(e_{3}\mu_{b}+e_{2}E_{p}^{b}+e'_{1}\epsilon_{p+q}^{e'}-e_{1}\epsilon_{q}^{e})]-1}{e_{3}\mu_{b}+e_{2}E_{p}^{b}+e'_{1}\epsilon_{p+q}^{e'}-e_{1}\epsilon_{q}^{e}}\nonumber \\
 &  & \times f_{B}(e_{3}\mu_{b}+e_{2}E_{p}^{b})f_{F}(-e_{1}\epsilon_{q}^{e})f_{F}(e'_{1}\epsilon_{p+q}^{e'})\frac{e_{2}}{2E_{p}^{b}}\cdot\frac{\epsilon_{p+q}^{e'}-e'_{1}\mu+e'e'_{1}E_{p+q}}{2\epsilon_{p+q}^{e'}}\cdot\frac{\epsilon_{q}^{e}+e_{1}\mu-ee_{1}E_{q}}{2\epsilon_{q}^{e}},\label{eq:2pi-form}\end{eqnarray}
where we have used \begin{eqnarray}
\Pi_{0}(\tau,p) & = & T\sum_{n}\Pi_{0}(p_{0},p)e^{-p_{0}\tau}\nonumber \\
 &  & =-8g^{2}\sum_{e,e',e_{1},e'_{1}}\int\frac{d^{3}q}{(2\pi)^{3}}\mathrm{Tr}[\Lambda_{q}^{e}\Lambda_{p+q}^{e'}]e^{(e'_{1}\epsilon_{p+q}^{e'}-e_{1}\epsilon_{q}^{e})(\beta-\tau)}\nonumber \\
 &  & \times f_{F}(-e_{1}\epsilon_{q}^{e})f_{F}(e'_{1}\epsilon_{p+q}^{e'})\frac{\epsilon_{p+q}^{e'}-e'_{1}\mu+e'e'_{1}E_{p+q}}{2\epsilon_{p+q}^{e'}}\cdot\frac{\epsilon_{q}^{e}+e_{1}\mu-ee_{1}E_{q}}{2\epsilon_{q}^{e}},\end{eqnarray}
and \begin{eqnarray}
D_{RR}(\tau,p) & = & \sum_{e_{2},e_{3}}\frac{e_{2}}{4E_{p}^{b}}f_{B}(e_{3}\mu_{b}+e_{2}E_{p}^{b})e^{(e_{3}\mu_{b}+e_{2}E_{p}^{b})\tau}.\end{eqnarray}

\end{document}